\documentclass[a4paper,american,aps,amsfonts,amsmath,prd,tightenlines, nofootinbib]{revtex4}
\usepackage[T1]{fontenc}
\usepackage[latin1]{inputenc}
\usepackage{graphicx}
\makeatletter

\usepackage{epsf}

\def\be{\begin{equation}}
\def\ee{\end{equation}}
\def\beq{\begin{equation}}
\def\eeq{\end{equation}}
\def\bea{\begin{eqnarray}}
\def\eea{\end{eqnarray}}
\def\bml{\begin{subequations}}
\def\blea{\bml\begin{eqnarray}}
\def\elea{\end{eqnarray}\end{subequations}}

\usepackage{babel}
\makeatother
\begin{document}

\title{Non-Gaussianity of the distribution tails in CMB}

\author{Vitaly Vanchurin}
\email{vitaly.vanchurin@physik.uni-muenchen.de}

\affiliation{Arnold-Sommerfeld-Center for Theoretical Physics, Department f\"{u}r
Physik, Ludwig-Maximilians-Universit\"{a}t M\"{u}nchen, Theresienstr.
37, D-80333, Munich, Germany}

\begin{abstract}
We analyze statistical properties of the separate multipole moments of the CMB temperature maps and find that the distribution tails are slightly non-Gaussian. Moreover, the deviation from Gaussianity peaks sharply at around $l\sim45\pm10$. If the detected non-Gaussianities should be attributed to the remaining foreground contamination from the galactic plane, an appearance of a similar peak in different frequency bands is unexpected. The presence of the peak was also confirmed by the analysis of weighted multipoles which is less susceptible to the galactic noise. A separate analysis of the multipoles with odd and even $l+m$ shows two opposite types of non-Gaussianities which cannot be explained with only galactic foregrounds. We argue that cosmic strings or correlated gravitational waves could lead to the observed effect.

\end{abstract}
\maketitle

\section{Introduction}

The CMB radiation is known to be Gaussian to a very high degree, which is in a good agreement with predictions of the simplest model of a single scalar field slow-roll inflation \cite{Linde, Maldacena}. It is also known that considerable non-Gaussianities could develop in the models of multi-field inflation \cite{BU, BKMR, Lyth,Weinberg}. Such scenarios usually predict non-vanishing three-point correlation functions which are often discussed in the literature in the context of WMAP data analysis \cite{GPW, WMAP, YW, Komatsu, Shellard}. At the same time the bispectrum is almost completely blind to non-Gaussianities from more complicated non-linear sources such as cosmic strings. Indeed, if the probability distribution of different multipole moments is nearly symmetric, then three-point correlation functions would not be very useful even if the distribution tails deviate significantly from a Gaussian. To detect such non-Gaussianities it seems necessary to use higher-point correlation functions (e.g. trispectrum). 

In this article, we discuss some methods which could be used to study the Gaussianity of the distribution tails. The  analysis of WMAP temperature maps was performed for different multipole moments in the range $20 < l < 100$, where the data are cleanest but the number of realizations of each random variable is sufficient to study the probability distributions of different multipoles without having to worry too much about cosmic variance.  Moreover, the separate $a_{lm}$'s are (nearly) independent random variables which simplifies the analysis significantly.

The paper is organized as follows. In Section II we describe the tests of Gaussianity which are employed to study the WMAP temperature maps. In Section III we discuss two different mechanisms involving cosmic strings and correlated gravitational waves, which could lead to the detected non-Gaussianities.

\section{Data analysis}

It is often assumed that separate multipole moments of the CMB sky are Gaussian random variables. If so, then the WMAP temperature maps are completely specified by only a single function, the angular power spectrum 
\begin{equation}
C_{l}\equiv\left\langle \left|a_{lm}\right|^{2}\right\rangle .\label{eq:power_spectrum}\end{equation}
Indeed, this would be true if the CMB temperature fluctuations were produced by one or few uncorrelated Gaussian processes. Then for a fixed (and large enough) $l$ the values of $a_{lm}$'s must be described fairly well by a two dimensional Gaussian distribution with variance $\sigma_{x}=\sigma_{y}=\frac{1}{\sqrt{2}}C_{l}$. Generally, we expect all of $l$ random complex variables\footnote{For the temperature fluctuations to remain real the condition $a_{lm}=a_{l-m}^{*}$ must be satisfied. This means that strictly speaking we are dealing with $2l+1$ random real numbers. One could get twice as much information by testing Gaussianity of the components of complex amplitudes separately, but to avoid unnecessary complications we restrict our analysis to multipole moments  with $m>0$.}  on each angular scale $l$ to be distributed according to the Gaussian probability density
\begin{equation}
p_l\left(a_{lm}\right)=\frac{1}{C_{l}\pi}e^{-\frac{\left|a_{lm}\right|^{2}}{C_{l}}}\label{eq:gaussian}\end{equation}

\begin{figure}
  \resizebox{\textwidth}{!}{\includegraphics{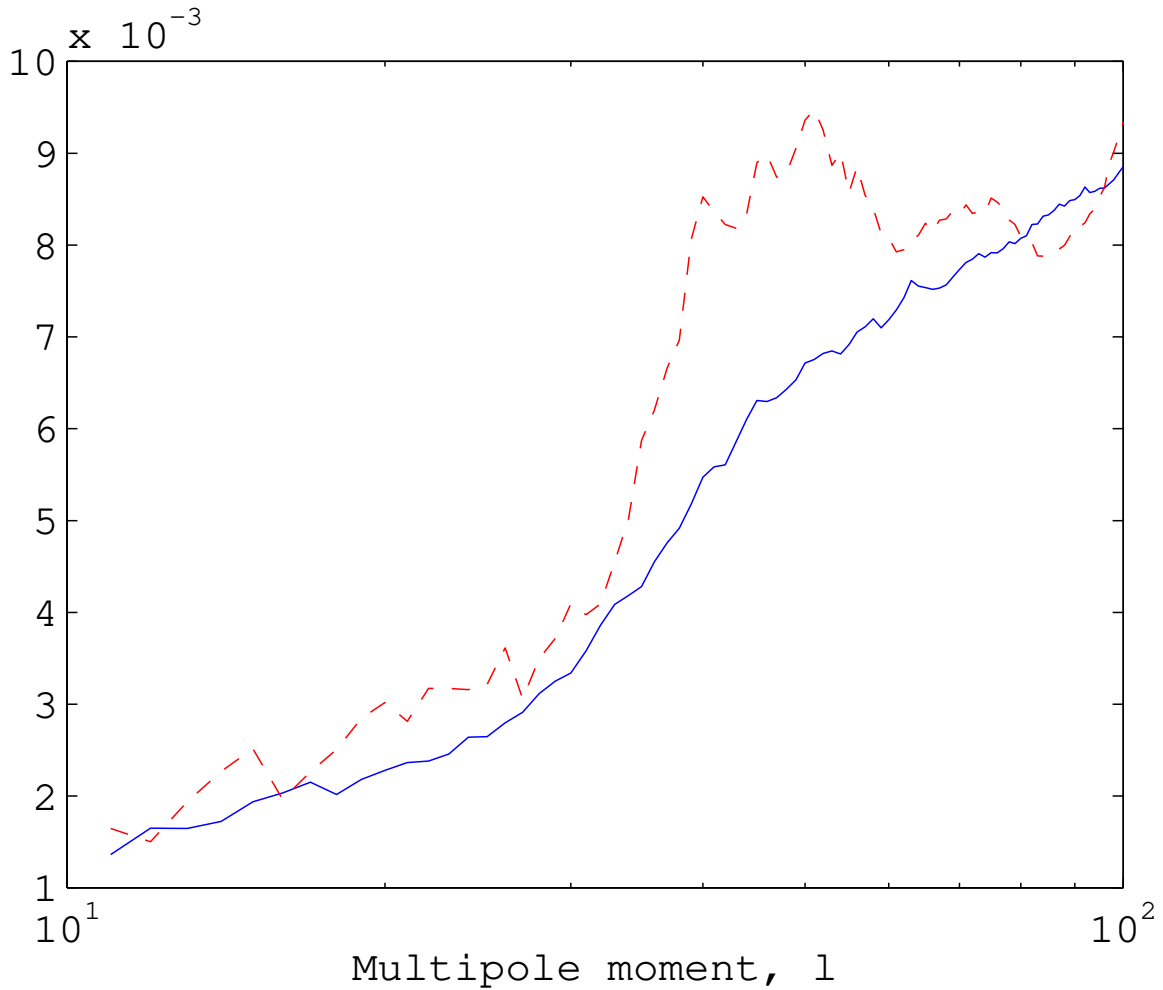}}
  \caption{$l(l+1) C_l/(2\pi)$ (solid blue) and $l(l+1) C^1_l/(2\pi)$ (dashed red) generated from foreground reduced WMAP full sky maps of W-band.\label{fig:smoothedw} } 
\end{figure}

\begin{figure}
  \resizebox{\textwidth}{!}{\includegraphics{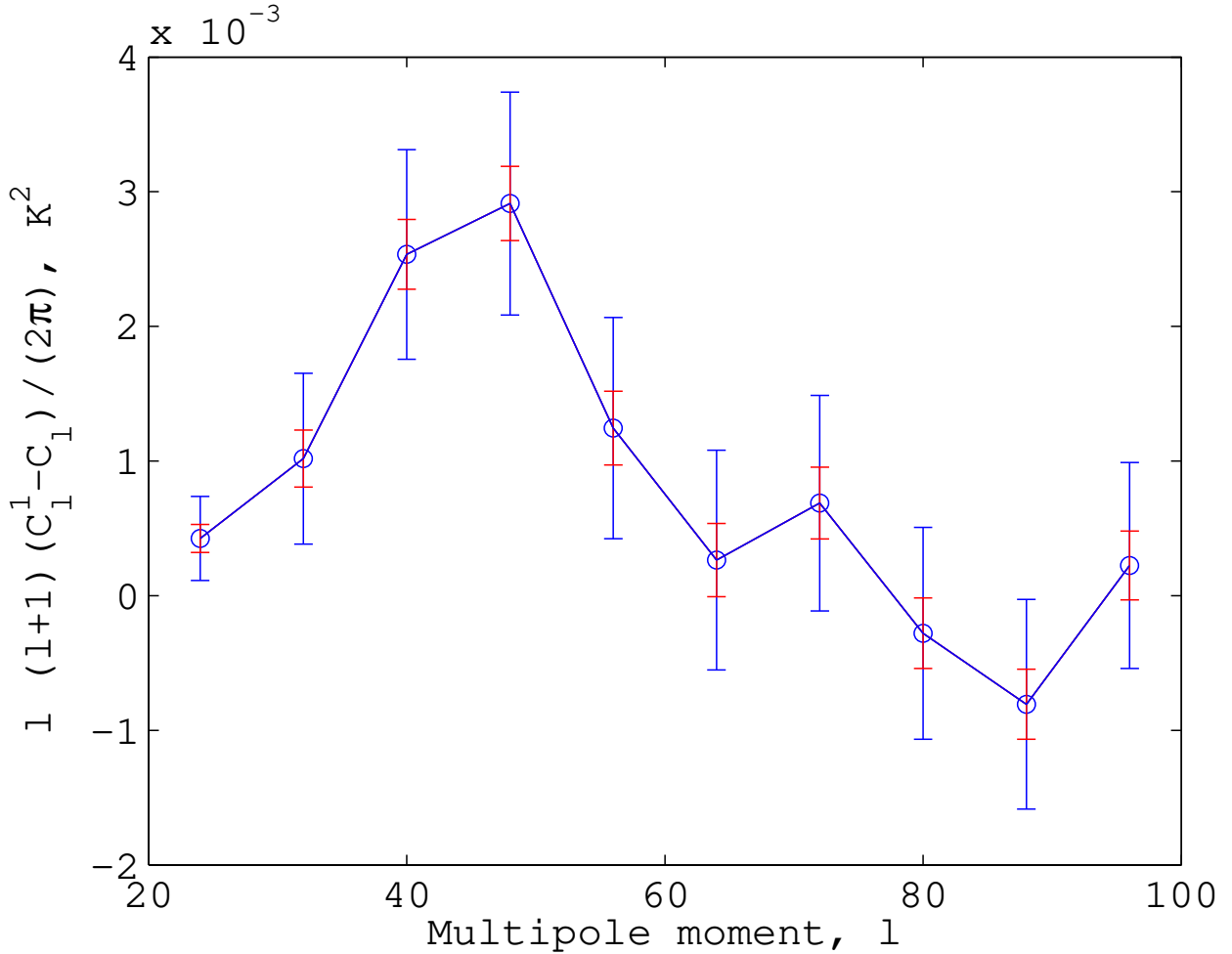}}
  \caption{Non-Gaussianity of foreground reduced WMAP full sky maps of W-band.   \label{fig:wband} } 
\end{figure}

\begin{figure}
  \resizebox{\textwidth}{!}{\includegraphics{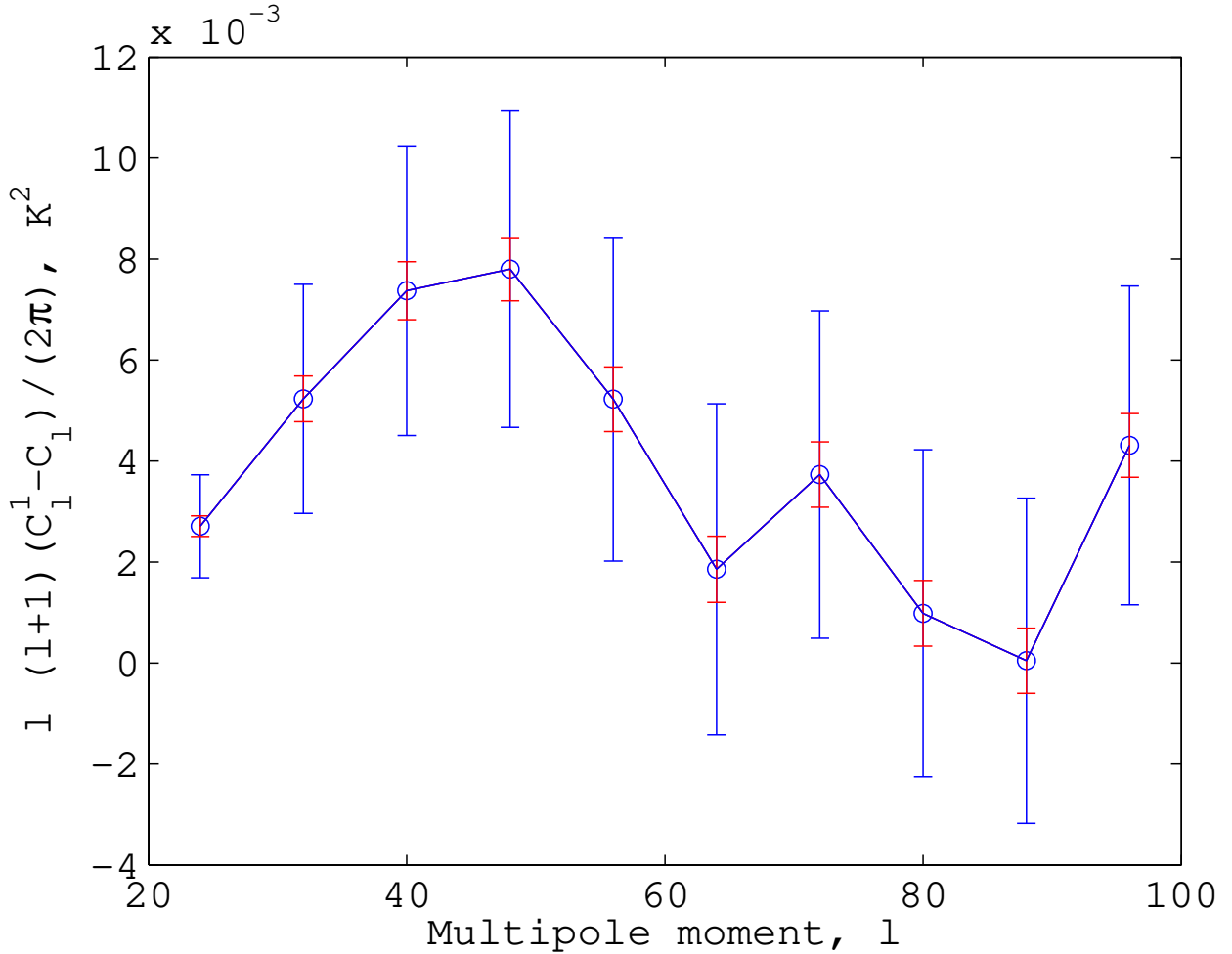}}
  \caption{Non-Gaussianity of foreground reduced WMAP full sky maps of V-band. \label{fig:vband} } 
\end{figure}

To evaluate the Gaussianity of an arbitrary distribution one usually looks at different statistical moments, which are quite noisy for a small number of data points. Since we are mainly interested in a statistics of the distribution tails $\left|a_{lm}\right|^{2}>\alpha C_{l}$
it will be convenient to define\footnote{More generally, one could be interested to test Gaussianity in an arbitrary range $\alpha C_{l}<\left|a_{lm}\right|^{2}<\beta C_{l}$ which is easily derived from $C_l^\alpha$ and $C_l^\beta$.}
\begin{equation}
C_{l}^{\alpha}\equiv \frac{e^{\alpha}}{1+\alpha}\int_{\sqrt{\alpha C_{l}}}^{\infty} r^2 p_l(r,\theta)\, 2\pi r\,dr  \label{eq:tails_spectrum}\end{equation}
where we switched to polar coordinates ($a_{lm}=r e^{i \theta}$) and the normalization was chosen such that  $C_{l}^{\alpha}=C_{l}$ for a Gaussian (see Eq. (\ref{eq:gaussian})).  On Fig. (\ref{fig:smoothedw})   we plot the smoothed $C_l$ and $C_l^\alpha$ generated from the foreground reduced WMAP full sky maps of W-band\footnote{The foreground reduced maps were obtained from the Legacy Archive for Microwave Background Data Analysis website and the HEALPix C++ routines were used to generate the corresponding $a_{lm}$'s.} (the frequency which is known to have the lowest foreground contamination). We choose a value of $\alpha=1$ such that the presence of a non-Gaussianity peak is most pronounced and the results have a sufficient statistical significance. One could clearly see an anomalous peak of the red dashed line in the range $35 \lesssim l \lesssim 55$. The feature can be better illustrated by binning and plotting only the deviations from Gaussianity. On Figs. (\ref{fig:wband}) and (\ref{fig:vband}) 
we plot $C_{l}^{\alpha}-C_{l}$ generated from two different frequency maps and see a very similar behavior in W- and V-bands. This indicates that unknown foregrounds are not likely to explain the presence of the non-Gaussianity peak.

To understand the statistical significance of the result we first calculate an expected number of data points in the analyzed range at a given angular scale:
\begin{equation}
N(l) = l \int_{\sqrt{\alpha C_{l}}}^{\infty} p_l(r)\, 2\pi r\, dr  =  l e^{-\alpha}. \label{eq:data_points}\end{equation}
Then the cosmic variance in the values of $C_l^\alpha$ with $N(l)$ data points\footnote{We neglect possible variations in the number of data points which is justified for $l>20$ and $\alpha\leq1$.} is given by 
\begin{equation}
\delta(l)  \approx \frac{1}{(1+\alpha)}  \frac{C_l}{\sqrt{N(l)}} 
\label{eq:variance}\end{equation}
which is approximately equal to $\delta(l) \approx 0.8\frac{C_l}{\sqrt{l}}$ for $\alpha=1$. The small red error bars plotted on Figs. (\ref{fig:wband}) and (\ref{fig:vband}) represent this variance with binning taken into account. That would be all if we had a clean data set, but the foreground reduced maps contain very high contamination form the galactic plane. The superposition of the galactic noise and CMB radiation increases significantly the total variance, which could be obtained from the observed variations at the angular scales away from the peak. The large blue error bars on Figs. (\ref{fig:wband}) and (\ref{fig:vband}) give a rough estimate of these uncertainties.

\begin{figure}
  \resizebox{\textwidth}{!}{\includegraphics{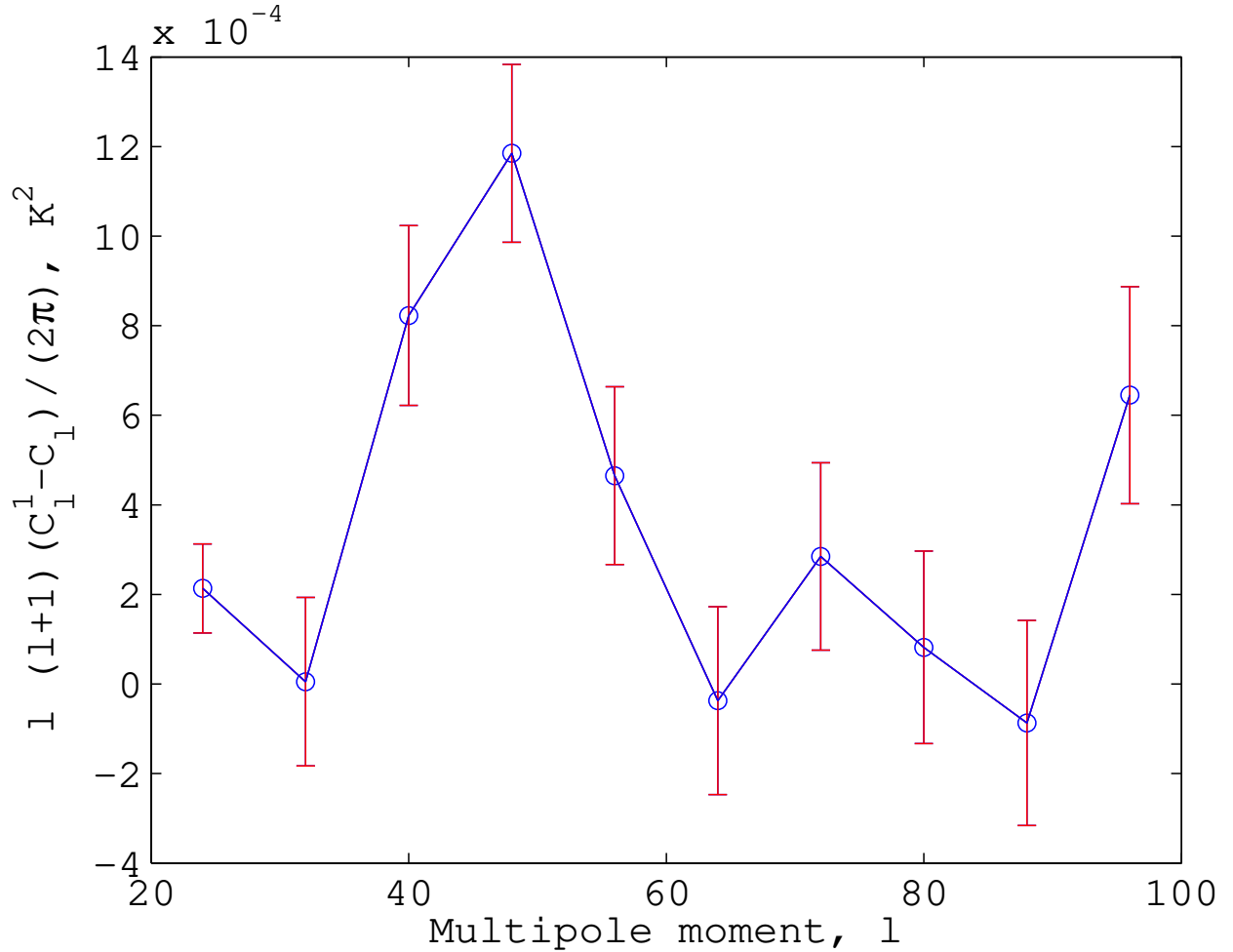}}
  \caption{Non-Gaussianity of CMB estimated by multipoles weighting of W-band map.  \label{fig:weightedw} } 
\end{figure}

To get a better insight, we reanalyze  the data by penalizing the multipole moments which are most susceptible to the galactic noise. Clearly, the spherical harmonics 
\begin{equation}
Y_{lm}\propto e^{i m \varphi} P_l^m(cos(\theta)),
\label{eq:shp_harmonmics}\end{equation}
with larger $m$ (corresponding to larger relative amplitudes of Legendre functions $P_l^m$ near equator and more frequent oscillations of $Y_{lm}$ along the azimuthal angle $\varphi$) pick up a stronger signal from the galaxy. To take into account unequal contamination of the multipoles we introduce weighting which increases the significance of  relatively clean modes, 
\begin{equation}
w(a_{lm}) = m^{-\beta},
\label{eq:weighting}\end{equation}
where $\beta>0$. On Fig. (\ref{fig:weightedw})   we plot non-Gaussianity of W-band map with weighting parameter $\beta=1$ and error bars described by Eq. (\ref{eq:variance}). One can see that the statistical variations associated with the galactic plane are greatly reduced, but the anomalous peak is still present. It follows that the probability for such an excessive non-Gaussianity to be present in a range from $l\approx36$ to $l\approx52$ is about $P\approx 0.000001$.\footnote{This result must be taken with caution, since the foreground reduced maps are highly contaminated at the galactic plane from where surprising non-Gaussianities might arise.} 

The analysis of the WMAP ILC\footnote{It is well known that the WMAP ILC map is not reliable at small angular  scales ($l\gtrsim 20$).} and Harmonic ILC\footnote{We are grateful to Pavel Naselsky for providing us with the Harmonic ILC temperature map  \cite{Naselsky}.} maps did not show any significant deviations from Gaussianity. Although, the linear combination techniques are known to introduce unphysical correlations between separate multipole moments, it remains to be seen whether this effect is significant enough to eliminate a non-Gaussian peak which is observed independently in different frequency bands.

\section{Discussion}

A preliminary analysis of the WMAP foreground reduced maps revealed some unexpected non-Gaussianities around $l\sim 45\pm10$. The standard inflation does not predict such effects, but one might always attribute them to unknown astrophysical sources. It is quite likely that foreground contamination of the galactic plane would modify the tails of distributions, but it is not very likely that the same non-Gaussian feature is present at exactly the same location in different frequency bands\footnote{The Q-band foreground reduced map contains a very similar non-Gaussianity peak at the same angular scales, but this frequency is known to be most contaminated by foregrounds, and as a result the analysis of the Q-band map was not conclusive.} unless the feature has a cosmological origin. We would also like to note that the anomaly appears approximately where an outlying point of the WMAP Collaboration published power spectrum is located \cite{WMAP}. 

To check our results we analyzed further the W-band map, which contains the lowest contamination, but nevertheless quite noisy at the galactic plane.  The main idea was to introduce unequal weighting of the multipole moments contaminated by foregrounds. The weighting technique allowed us to significantly reduce the uncertainties (See Fig. (\ref{fig:weightedw})) and to demonstrate that the observed non-Gaussianity is not likely to come from only galactic plane. Let us now describe two different mechanisms which could naturally lead to changing non-Gaussian tails at the angular scale of the anomaly. 

During inflationary stage the presence of non-scalar background fields (e.g. vector inflation \cite{GMV}) could lead to a mixture of perturbations already at the linear order  \cite{GMV-grav,GMV-pert}. Higher order corrections enhance the mixture (even after inflation \cite{stein}) leading to non-trivial correlations between Gaussian random variables (e.g. scalar and tensor components). Both effects combined could lead to appearance of non-Gaussian distribution tails. If we assume that primordial gravitational waves contribute a significant power to the CMB temperature fluctuations and that the tensor modes became correlated with scalar modes, then changes in the behavior of one of the components could produce changes in non-Gaussianity. In fact the place where an anomalous peak appears is exactly where the tensor contribution to the temperature power spectrum should start a very rapid decline due to decaying evolution inside the horizon \cite{Mukhanov}. If this interpretation is correct, then one would expect the non-Gaussianity to change significantly close to the angular scales of $l\sim 45$.

Although correlated gravitational waves could produce significant changes in non-Gaussianity, it is hard to argue that such scenarios are typical. In contrast, the production of cosmic strings is a generic feature of many models involving symmetry breaking phase transitions \cite{VilenkinShellard}. Non-linear interaction of the CMB photons with a network of cosmic strings could easily generate non-Gaussian fluctuations in the temperature. Even mild discontinuities of the CMB temperature caused by cosmic strings are translated directly into appearances of non-Gaussian distribution tails \cite{FRSB, Sasaki}. The angular scales of $l \sim 200$ correspond to the horizon size at the last scattering surface where the largest power from cosmic strings \cite{LS, BHKU} and also from inflationary perturbation \cite{Mukhanov} is expected. Thus, the relevant question is at which $l$ the relative contribution from cosmic strings to the overall power is at its highest value. The answer to this highly non-trivial problem depends crucially on the evolution of the cosmic string network \cite{BB, AS, VOV} and in particular on the distribution of the cosmic string loops \cite{VOV2, MS, RSB, OV, Vanchurin}. It is quite natural to expect that at $l\sim 45$ where the power spectrum from inflationary perturbation just starts to rise, the relative contribution from cosmic strings is the highest. As a result, one should see the largest deviation from Gaussianity to be somewhat close to  where the non-Gaussianity peak is in fact observed.

Both mechanisms (cosmic strings or correlated gravitational waves) could in principle be employed to explain the anomaly, but before we seriously consider such scenarios it is important to fully understand the effect. This could only be achieved by performing a lot more sophisticated tests at the angular scales: $35 \lesssim l \lesssim 55$. The detected non-Gaussianity could not have been seen from the usual bispectrum, but should in principle be observable through the trispectrum analysis (or through the forth statistical moment). It would also be desirable to probe larger multipoles ($l\gtrsim100$) in order to favor one model or the other.  All of these issues deserve further investigation. 

\begin{acknowledgments}

The authors are grateful to Richard Bond,  Alexey Golovnev, Stefan Hofmann, Slava Mukhanov, Pavel Naselsky, Levon Pogosian, Amir Said, Mikhail Sazhin, Dominik Schwarz, Alex Vilenkin and Sergei Winitzki for very useful discussions. This work was supported in part by TRR 33 {}``The Dark Universe'' and the Cluster of Excellence EXC 153 {}``Origin and Structure of the Universe''. 
\end{acknowledgments}

\section*{Note Added}

After the first version of the paper was submitted we learned from Pavel Naselsky another way to reduces the possible contamination from the galactic plane. Apparently the multipoles with odd $l+m$ whose spherical harmonics vanish at the equator are less affected by the galactic noise. We had preformed a similar analysis with only odd $l+m$ multipoles and discovered an opposite non-Gaussiniaty at the angular scales of the anomaly. Instead of having too many outlier we have to few, but the non-Gaussianity is still present at around $l \sim 45 \pm 10$. It is hard to imagine how the galactic noise alone could act in opposite direction for odd and even $l+m$ multipoles in only a narrow range of scales. In fact, similar non-Gaussianity was recently reported by another group \cite{Wetterich}. The authors looked at the abundances of cold and hot spots in CMB and found a lack of power on the angular scales of several degrees.

\end{document}